\documentstyle[preprint,aps,amssymb,amscd,epsfig]{revtex}
\begin{document}


\def\Lb{\Lambda_b}
\def\Lce{\Lambda_{c1}}
\def\Lca{\Lambda_{c1}^{\frac{1}{2}}}
\def\Lcb{\Lambda_{c1}^{\frac{3}{2}}}
\def\vslash{v\hspace{-1.8mm}/}
\def\vpslash{v^\prime\hspace{-2.8mm}/}
\def\Pa{\frac{1+\vslash}{2}}
\def\Pb{\frac{1+\vpslash}{2}}
\def\LQCD{\Lambda_{\rm QCD}}
\def\Pc{\frac{1-\vslash}{2}}
\def\Pd{\frac{1-\vpslash}{2}}
\def\Dl{{\overleftarrow D}}
\def\Dr{{\overrightarrow D}}
\def\Dsl{{\overleftarrow D}\hspace{-3.0mm}/}
\def\Dsr{{\overrightarrow D}\hspace{-3.0mm}/}
\def\Ds{D\hspace{-3.0mm}/}
\def\hbarc{{\bar h}^{(c)}_{v^\prime}}
\def\hb{h^{(b)}_v}
\def\qq{\langle{\bar q}q\rangle}
\def\qGq{\langle{\bar q}g\sigma\cdot G q\rangle}
\def\GG{\langle \alpha_s GG\rangle}


\title{Subleading Isgur-Wise Function of $\Lb\to\Lce$ using QCD sum rules}
\author{
Jong-Phil Lee\footnote{e-mail: jplee@phya.yonsei.ac.kr} and
Gye T.\ Park\footnote{e-mail :gtpark@phya.yonsei.ac.kr}}
\address{Department of Physics and IPAP, Yonsei University, Seoul, 120-749, Korea}
\maketitle


\begin{abstract}

Subleading Isgur-Wise form factor $\tau(v\cdot v')$ at ${\cal O}(1/m_Q)$ for
$\Lb\to\Lce^{1/2,3/2}$ weak transition
is calculated by using the QCD sum rules in the framework of the heavy quark
effective theory (HQET), where
$\Lambda^{1/2}_{c1}$ and $\Lambda^{3/2}_{c1}$ are the orbitally excited charmed 
baryon doublet with $J^P=(1^-/2,3^-/2)$.  
We consider the subleading contributions from the weak current matching in the
HQET.
The interpolating currents with 
transverse covariant derivative are adopted for $\Lambda^{1/2}_{c1}$ and 
$\Lambda^{3/2}_{c1}$ in the analysis.  
The slope parameter $\rho^2$ in linear 
approximation of $\tau$ is obtained to be $\rho^2=2.76$ and 
the interception to be $\tau(1)=-1.27$ GeV. 

\end{abstract}


\pagebreak

\section{Introduction}

The ground state bottom baryon $\Lambda_b$ weak decays \cite{PDG} provide a 
testing ground for the standard model (SM).  
They reveal some important features of the physics of bottom quark.  
The experimental data on these decays have been accumulated to wait for 
reliable theoretical calculations.  
With the discovery of the orbitally excited charmed baryons $\Lambda_c(2593)$ 
and $\Lambda_c(2625)$ \cite{CLEO}, it would be of great interest for one to 
investigate the $\Lambda_b$ semileptonic decays into these baryons.  
\par
From the phenomenological point of view, 
these semileptonic transitions are interesting since in 
principle they may account for a sizeable fraction of the inclusive 
semileptonic rate of $\Lambda_b$ decay.
In addition, the properties of excited  baryons have attracted attention in 
recent years.  
Investigation on them will extend our ability in the application of QCD.  
It can also help us foresee any other excited heavy baryons that have not been 
discovered yet.  
\par
The heavy quark symmetry \cite{HQET} is a useful tool to classify the hadronic 
spectroscopy containing a heavy quark $Q$.  
In the infinite mass limit, the 
spin and parity of the heavy quark and that of the light degrees of freedom are 
separately conserved. Coupling the spin of light degrees of freedom $j_\ell$ 
with the spin of heavy quark $s_Q=1/2$ yields a doublet  with total spin 
$J=j_\ell\pm 1/2$ (or a singlet if $j_\ell=0$). This classification can be 
applied to the $\Lambda_Q$-type baryons. For the charmed baryons the ground 
state $\Lambda_c$ contains light degrees of freedom with spin-parity 
$j_\ell^{P}=0^+$, being a singlet. The excited states with $j_\ell^P=1^-$ are 
spin symmetry doublet with $J^P$($1^-/2$,$3^-/2$). 
The lowest states of such 
excited charmed states, $\Lambda^{1/2}_{c1}$ and $\Lambda^{3/2}_{c1}$, have 
been observed to be identified with $\Lambda_c(2593)$ and $\Lambda_c(2625)$ 
respectively \cite{CLEO}.
\par
However, the difficulties in the SM calculations are mainly due to the poor 
understanding of the nonperturbative aspects of the strong interaction (QCD).  
The heavy 
quark effective theory (HQET) based on the heavy quark symmetry provides a 
model-independent method for analyzing heavy hadrons containing a single heavy 
quark \cite{HQET}.  It allows us to expand the physical quantity in powers of 
$1/m_Q$ systematically, where $m_Q$ is the heavy quark mass.  
Within this 
framework, the classification of the $\Lambda_b$ exclusive weak decay form 
factors has been greatly simplified.  
The decays such as
$\Lambda_b\to\Lambda_c l{\bar\nu}$ \cite{IW}, 
$\Lambda_b\to\Sigma_c^{(*)}l{\bar\nu}$ \cite{Mannel},
$\Lambda_b\to\Sigma_c^{(*)} \pi l{\bar\nu}$ \cite{Cho}, 
$\Lambda_b\to p(\Lambda)$ \cite{Qiao} have been studied.
\par
To obtain detailed predictions for the hadrons, at this point, 
some nonperturbative QCD methods are also required.  
We have adopted QCD sum rules \cite{sumrule0} in this work.
QCD sum rule is a powerful nonperturbative method based on QCD.
It takes into account the nontrivial QCD vacuum which is parametrized by various
vacuum condensates in order to describe the nonperturbative nature.  
In QCD sum rule, hadronic observables can be calculated by evaluating two- or 
three-point correlation functions.  
The hadronic currents for constructing the correlation 
functions are expressed by the interpolating fields.  
In describing the excited heavy baryons, transverse covariant derivative is 
included in the interpolating field.
The static properties of 
$\Lambda_b$ and $\Lambda_{c1}$ ($\Lambda_{c1}$ denotes the generic 
$j_\ell^{P}=1^-$ charmed state) have been studied with QCD sum rules in the 
HQET in Ref.\ \cite{sumrule} and Ref.\ \cite{PLB476,Zhu}, respectively.  
Recently, the leading order Isgur-Wise (IW) function is also calculated in the 
HQET QCD sum rule in Ref.\ \cite{PLB502}.
\par
In $\Lb\to\Lce$ decay, $1/m_Q$ corrections are very important.
At the heavy quark limit of $m_Q\to\infty$, the transition matrix elements 
should vanish at zero recoil since the light degrees of freedom change their 
configurations.
Nonvanishing contribution to, say, ${\cal B}(\Lb\to\Lce\ell{\bar\nu})$ at zero
recoil appears at $1/m_Q$ order.
Since both $\Lb$ and $\Lce$ are heavy enough, 
the behavior of the matrix elements near the zero recoil is very important.
That explains why people pay attention to the next-to-leading order (NLO) 
contributions.
The same situation occurs in heavy mesons.
As for $B\to D_1(D_2^*)\ell {\bar\nu}$ decay, 
leading and subleading Isgur-Wise (IW) functions have been computed using QCD 
sum rule in Ref.\
\cite{Ligeti,Colangelo,Dai,Huang,Wang}.
They showed that the branching ratio is enhanced considerably when the 
subleading contributions are included.
\par
In HQET, $1/m_Q$ corrections appear in a two-fold way.
At the Lagrangian level, subleading terms are summarized in $\lambda_1$ and
$\lambda_2$. 
$\lambda_1$ parametrizes the kinetic term of higher 
derivative, while $\lambda_2$ represents the chromomagnetic interaction which
explicitly breaks the heavy quark spin symmetry.
At the current level, $1/m_Q$ corrections come from the small portion of the
heavy quark fields which correspond to the virtual motion of the heavy quark.
In this work, the subleading IW function from the latter case, i.e., at the
current level, is analyzed in the HQET QCD sum rules.
\par
In Sec.\ II, the weak transition matrix elements are parametrized by the 
leading and subleading IW functions.
By evaluating the three-point correlation function, we give the subleading
IW function in Sec.\ III.
We present, in Sec. IV, the numerical analysis and discussions.
The summary is given in Sec.\ V.

\section{Weak Transition Matrix Elements and the Subleading Isgur-Wise 
Functions}

The weak transition matrix elements for $\Lb\to\Lce$ are parametrized by the
14-form factors as
\begin{mathletters}\label{formfactor}
\begin{eqnarray}
\frac{\langle\Lca(v^\prime,s^\prime)|V_\mu|\Lambda_b(v,s)\rangle}
{\sqrt{4M_{\Lambda_{c1}(1/2)}M_{\Lambda_b}}}&=&
{\bar u}_{\Lambda_{c1}}(v^\prime,s^\prime)
 \Big[F_1\gamma_\mu+F_2 v_\mu+F_3 v^\prime_\mu\Big]
\gamma_5 u_{\Lambda_b}(v,s)~,\\
\frac{\langle\Lca(v^\prime,s^\prime)|A_\mu|\Lambda_b(v,s)\rangle}
{\sqrt{4M_{\Lambda_{c1}(1/2)}M_{\Lambda_b}}}&=&
{\bar u}_{\Lambda_{c1}}(v^\prime,s^\prime)
 \Big[G_1\gamma_\mu+G_2 v_\mu+G_3 v^\prime_\mu\Big]
u_{\Lambda_b}(v,s)~,\\
\frac{\langle\Lcb(v^\prime,s^\prime)|V_\mu|\Lambda_b(v,s)\rangle}
{\sqrt{4M_{\Lambda_{c1}(3/2)}M_{\Lambda_b}}}&=&
{\bar u}^\alpha_{\Lambda_{c1}}(v^\prime,s^\prime)\Big[
 v_\alpha(K_1\gamma_\mu+K_2 v_\mu+K_3 v^\prime_\mu)+K_4 g_{\alpha\mu}\Big]
u_{\Lambda_b}(v,s)~,\\
\frac{\langle\Lcb(v^\prime,s^\prime)|A_\mu|\Lambda_b(v,s)\rangle}
{\sqrt{4M_{\Lambda_{c1}(3/2)}M_{\Lambda_b}}}&=&
{\bar u}^\alpha_{\Lambda_{c1}}(v^\prime,s^\prime)\Big[
v_\alpha(N_1\gamma_\mu+N_2 v_\mu+N_3 v^\prime_\mu)+N_4 g_{\alpha\mu}\Big]
\gamma_5 u_{\Lambda_b}(v,s)~,
\end{eqnarray}
\end{mathletters}
where $v(v')$ and $s(s')$ are the four-velocity and spin of 
$\Lambda_b(\Lambda_{c1})$, respectively.  And the form factors $F_i$, $G_i$, 
$K_i$ and $N_i$ are functions of $y\equiv v\cdot v'$.  
In the limit of $m_Q\to\infty$,
all the form factors are related to one independent universal form factor
$\xi(y)$ called Isgur-Wise (IW) function.
A convenient way to evaluate 
hadronic matrix elements is by introducing interpolating fields in HQET 
developed in Ref.~\cite{Falk} to parametrize the matrix elements in 
Eqs. (\ref{formfactor}).  With the aid of this method the matrix element can be 
written as \cite{Leibovich}
\begin{equation}
{\bar c}\Gamma b={\bar h}^{(c)}_{v^\prime}\Gamma h^{(b)}_v
=\xi(y)v_\alpha{\bar\psi}^\alpha_{v^\prime}\Gamma\psi_v~
\label{current}
\end{equation}
at leading order in $1/m_Q$ and $\alpha_s$, where $\Gamma$ is any collection of 
$\gamma$-matrices. The ground state field, $\psi_v$, destroys the $\Lambda_b$ 
baryon with four-velocity $v$; the spinor field $\psi^\alpha_v$ is given by
\begin{equation}
\psi^\alpha_v=\psi^{3/2\alpha}_v
 +\frac{1}{\sqrt{3}}(\gamma^\alpha+v^\alpha)\gamma_5\psi^{1/2}_v~,
\label{spinor3}
\end{equation}
where $\psi^{1/2}_v$ is the ordinary Dirac spinor and $\psi^{3/2\alpha}_v$ is 
the spin 3/2 Rarita-Schwinger spinor, they destroy $\Lambda^{1/2}_{c1}$ and 
$\Lambda^{3/2}_{c1}$ baryons with four-velocity $v$, respectively.
To be explicit,
\begin{eqnarray}
F_1&=&\frac{1}{\sqrt{3}}(y-1)~\xi(y)~,~~~
G_1=\frac{1}{\sqrt{3}}(y+1)~\xi(y)~,\nonumber\\
F_2&=&G_2=-\frac{2}{\sqrt{3}}~\xi(y)~,~~~~~
K_1=N_1=\xi(y)~,\nonumber\\
&&~~~~~~~~~~~~~~~~~({\rm others})=0~.
\label{leadingform}
\end{eqnarray}
In general, the IW form factor is a decreasing function of the four velocity
transfer $y$.  Since the kinematically allowed region of $y$ for heavy to 
heavy transition is very narrow around unity,
\begin{equation}
1\le y\le 
\frac{M_{\Lambda_b}^2+M_{\Lambda_{c1}}^2}{2M_{\Lambda_b}M_{\Lambda_{c1}}}
\simeq 1.3~,
\end{equation}
and hence it is convenient to approximate the IW function linearly as
\begin{equation}
\xi(y)=\xi(1)(1-\rho_\xi^2(y-1))~,
\label{rho}
\end{equation}
where $\rho_\xi^2$ is the slope parameter which characterizes the shape of the
leading IW function.
\par
The $\LQCD/m_Q$ corrections come in two ways.
One is from the subleading Lagrangian of the HQET while the other comes from the
small portion of the heavy quark field to modify the effective currents.
We only consider the latter case here.
\par
Including $\LQCD/m_b$ and $\LQCD/m_c$, the weak current is given by
\begin{equation}
{\bar c}\Gamma b=\hbarc\Bigg(\Gamma-\frac{i}{2m_c}\Dsl\Gamma
 +\frac{i}{2m_b}\Gamma\Dsr\Bigg)\hb~.
\end{equation}
Keeping the Lorentz structure, the subleading terms are expanded in general as
\begin{eqnarray}
\hbarc i\Dsl\Gamma\hb&=&{\bar\psi}^\alpha_{v^\prime}
 (\tau_1^{(c)}v_\alpha\vslash+\tau_2^{(c)}v_\alpha\vpslash
 +\tau_3^{(c)}\gamma_\alpha)\Gamma\Lambda_v~,\nonumber\\
\hbarc\Gamma i\Dsr \hb&=&{\bar\psi}^\alpha_{v^\prime}\Gamma
 (\tau_1^{(b)}v_\alpha\vslash+\tau_2^{(b)}v_\alpha\vpslash
 +\tau_3^{(b)}\gamma_\alpha)\Lambda_v~,
\label{subleading}
\end{eqnarray}
where $\tau_i^{(Q)}$ are the subleading IW functions to be evaluated.
\par
The matrix elements of these currents modify Eq.\ (\ref{leadingform}) as
\begin{eqnarray}\label{FGKN}
\sqrt{3}F_1&=&(y-1)\xi-\epsilon_c\Big[
   (y-1)(-\tau_1^{(c)}+\tau_2^{(c)})+3\tau_3^{(c)}\Big]
 +\epsilon_b\Big[(y-1)(\tau_1^{(b)}-\tau_2^{(b)})-\tau_3^{(b)}\Big],\nonumber\\
\sqrt{3}F_2&=&-2\xi+\epsilon_c\Big[
   2y\tau_1^{(c)}+2\tau_2^{(c)}\Big]
 +\epsilon_b\Big[-2\tau_1^{(b)}+2\tau_2^{(b)}\Big]~,\nonumber\\
\sqrt{3}F_3&=&-2\epsilon_b\Big[(1+y)\tau_2^{(b)}+\tau_3^{(b)}\Big]~,\nonumber\\
\sqrt{3}G_1&=&(y+1)\xi-\epsilon_c\Big[
   (y+1)(\tau_1^{(c)}+\tau_2^{(c)})+3\tau_3^{(c)}\Big]
 +\epsilon_b\Big[(y+1)(\tau_1^{(b)}+\tau_2^{(b)})+\tau_3^{(b)}\Big]~,\nonumber\\
\sqrt{3}G_2&=&-2\xi+\epsilon_c\Big[2y\tau_1^{(c)}+2\tau_2^{(c)}\Big]
 -2\epsilon_b\Big[\tau_1^{(b)}+\tau_2^{(b)}\Big]~,\nonumber\\
\sqrt{3}G_3&=&2\epsilon_b\Big[(y-1)\tau_2^{(b)}+\tau_3^{(b)}\Big]~,\nonumber\\
K_1&=&\xi+\epsilon_c\Big[\tau_1^{(c)}-\tau_2^{(c)}\Big]
 +\epsilon_b\Big[\tau_1^{(b)}-\tau_2^{(b)}\Big]~,\nonumber\\
N_1&=&\xi-\epsilon_c\Big[\tau_1^{(c)}+\tau_2^{(c)}\Big]
 +\epsilon_b\Big[\tau_1^{(b)}+\tau_2^{(b)}\Big]~,\nonumber\\
K_2&=&N_2=-2\epsilon_c\tau_1^{(c)}~,\nonumber\\
K_3&=&-N_3=2\epsilon_b\tau_2^{(b)}~,\nonumber\\
K_4&=&-N_4=2\epsilon_b\tau_3^{(b)}~,
\label{NLOff}
\end{eqnarray}
where $\epsilon_Q\equiv 1/2m_Q$.
It is quite convenient to define
\begin{mathletters}
\label{Omega}
\begin{eqnarray}
\Omega_{\alpha\beta}^{(c\Gamma)}&\equiv&
(\gamma_\alpha+v^\prime_\alpha)\gamma_5\Bigg(\Pb\Bigg)\gamma_\beta\Gamma
 \Bigg(\Pa\Bigg)~,\\
\Omega_{\alpha\beta}^{(b\Gamma)}&\equiv&
(\gamma_\alpha+v^\prime_\alpha)\gamma_5\Bigg(\Pb\Bigg)\Gamma\gamma_\beta
 \Bigg(\Pa\Bigg)~.
\end{eqnarray}
\end{mathletters}
Possible contractions of $\Omega_{\alpha\beta}$ are listed in the Appendix.
From the Eqs.\ (\ref{spinor3}) and (\ref{subleading}),
Eq.\ (\ref{formfactor}) can be reexpressed in terms of $\tau_i^{(Q)}$ and
$\Omega_{\alpha\beta}$:
\begin{eqnarray}\label{ff2}
\frac{\langle\Lca(v^\prime,s^\prime)|\Gamma|\Lambda_b(v,s)\rangle}
{\sqrt{4M_{\Lambda_{c1}(1/2)}M_{\Lambda_b}}}
&=&\frac{1}{\sqrt{3}}
{\bar u}_{\Lambda_{c1}}(v^\prime,s^\prime)
 \Big[\xi v^\alpha v^{\prime\alpha}\Omega^{(c\Gamma)}_{\alpha\beta}
   -\epsilon_c\Big(\tau_1^{(c)}v^\alpha v^\beta
     +\tau_2^{(c)}v^\alpha v^{\prime\beta}
     +\tau_3^{(c)}g^{\alpha\beta}\Big)\Omega^{(c\Gamma)}_{\alpha\beta}
  \nonumber\\
  && +\epsilon_b\Big(\tau_1^{(b)}v^\alpha v^\beta
     +\tau_2^{(b)}v^\alpha v^{\prime\beta}
     +\tau_3^{(b)}g^{\alpha\beta}\Big)\Omega^{(b\Gamma)}_{\alpha\beta}
 \Big]u_{\Lambda_b}(v,s)~,\\
\end{eqnarray}
A similar expression can be obtained for the spin-3/2 final states 
\begin{eqnarray}
\frac{\langle\Lcb(v^\prime,s^\prime)|\Gamma|\Lambda_b(v,s)\rangle}
{\sqrt{4M_{\Lambda_{c1}(3/2)}M_{\Lambda_b}}}
&=&
{\bar u}^\alpha_{\Lambda_{c1}}(v^\prime,s^\prime)
 \Big[\xi v_\alpha \Gamma
   -\epsilon_c\Big(\tau_1^{(c)}v_\alpha v_\beta
     +\tau_2^{(c)}v_\alpha v^{\prime}_\beta
     +\tau_3^{(c)}g_{\alpha\beta}\Big)\gamma^\beta\Gamma\nonumber\\
  && +\epsilon_b\Big(\tau_1^{(b)}v_\alpha v_\beta
     +\tau_2^{(b)}v_\alpha v^{\prime}_\beta
     +\tau_3^{(b)}g_{\alpha\beta}\Big)\Gamma\gamma^\beta
 \Big]u_{\Lambda_b}(v,s)~.
\end{eqnarray}

\section{QCD sum rule evaluation}

As a starting point of QCD sum rule calculation, let us consider the 
interpolating field of heavy baryons.  
The heavy baryon current is generally expressed as 
\begin{equation}
j^v_{J,P}(x)=\epsilon_{ijk}
[q^{iT}(x)C\Gamma_{J,P}\tau q^j(x)]\Gamma^\prime_{J,P} h^k_v(x)~,
\end{equation}
where $i,~j,~k$ are the color indices, $C$ is the charge conjugation matrix, 
and $\tau$ is the isospin matrix while $q(x)$ is a light quark field.
$\Gamma_{J,P}$ and $\Gamma^\prime_{J,P}$ are some gamma matrices which 
describe the structure of the baryon with spin-parity $J^P$.  Usually $\Gamma$ 
and $\Gamma^\prime$ with least number of derivatives are used in the QCD sum 
rule method.  The sum rules then have better convergence in the high energy 
region and often have better stability.  For the ground state heavy baryon, we 
use $\Gamma_{1/2,+}=\gamma_5$, $\Gamma^\prime_{1/2,+}=1$.  In the previous 
work \cite{PLB476}, two kinds of interpolating fields are introduced to
represent the excited heavy baryon.  In this work, we find that only the 
interpolating field of transverse derivative is adequate for the analysis.
Nonderivative interpolating field results in a vanishing perturbative
contribution.  The choice of $\Gamma$ and $\Gamma^\prime$ with derivatives for 
the $\Lambda_{c1}^{1/2}$ and $\Lambda_{c1}^{3/2}$ is then 
\begin{eqnarray}
\Gamma_{1/2,-}&=&(a+b\vslash)\gamma_5~,~~~
\Gamma^\prime_{1/2,-}=\frac{i{\Dsl}_t}{M}\gamma_5~,\nonumber\\
\Gamma_{3/2,-}&=&(a+b\vslash)\gamma_5~,~~~
\Gamma^\prime_{3/2,-}=
 \frac{1}{3M}(i{\overleftarrow D}^\mu_t+i{\Dsl}_t\gamma^\mu_t)~,
\label{deriv}
\end{eqnarray}
where a transverse vector $A^\mu_t$ is defined to be 
$A^\mu_t\equiv A^\mu-v^\mu v\cdot A$, and $M$ in Eq. (\ref{deriv}) is some 
hadronic mass scale.  $a$, $b$ are arbitrary numbers between 0 and 1.
\par
The baryonic decay constants in the HQET are defined as follows,
\begin{mathletters}
\begin{eqnarray}
\langle 0|j^v_{1/2,+}|\Lambda_{b}\rangle &=&f_{\Lambda_b}\psi_v~,\\
\langle 0|j^{v}_{1/2,-}|\Lambda_{c1}^{1/2}\rangle &=&f_{1/2}
\psi^{1/2}_{v}~,\\
\langle 0|j^{v\mu}_{3/2,-}|\Lambda_{c1}^{3/2}\rangle &=&
\frac{1}{\sqrt{3}}f_{3/2}\psi^{3/2\mu}_{v}\;,
\end{eqnarray}
\end{mathletters}
where $f_{1/2}$ and $f_{3/2}$ are equivalent since $\Lambda_{c1}^{1/2}$ and 
$\Lambda_{c1}^{3/2}$ belong to the same doublet with $j_\ell^P=1^-$. The 
QCD sum rule calculations give \cite{sumrule} 
\begin{equation}
f_{\Lambda_b}^2 e^{-{\bar\Lambda}/T}=\frac{1}{20\pi^4}
\int_0^{\omega_c}d\omega \omega^5 e^{-\omega/T}
+\frac{1}{6}\langle{\bar q}q\rangle^2 e^{-m_0^2/8T^2}
+\frac{\langle\alpha_s GG\rangle}{32\pi^3}T^2~,
\end{equation}
and \cite{PLB476}
\begin{eqnarray}
M^2f_{1/2}^2 e^{-{\bar\Lambda}^\prime/T^\prime}&=&
\int_0^{\omega_c^\prime} d\omega \frac{3N_c!}{4\pi^4\cdot 7!}\omega^7
(24a^2+40b^2)e^{-\omega/T^\prime}+\frac{\langle\alpha_s GG\rangle}{32\pi^3}
T^{\prime 4}(-a^2+b^2)\nonumber\\
&&+\frac{N_c!}{2\pi^2}\Big[\langle{\bar q}q\rangle T^{\prime 5}(16ab)
 -\langle{\bar q}g\sigma\cdot G q\rangle T^{\prime 3}ab\Big]
-\frac{\langle{\bar q}g\sigma\cdot G q\rangle}{4\pi^2}T^{\prime 3}(3ab)~.
\end{eqnarray}
In the above equations, $T^{(\prime)}$ are the Borel parameters and
$\omega_c^{(\prime)}$ are the continuum thresholds, and $N_c=3$ is the
color number.  In the heavy quark limit, the mass parameters $\bar\Lambda$ 
and $\bar\Lambda^\prime$ are defined as
\begin{equation}
{\bar\Lambda}^\prime=M_{\Lambda_{Q1}}-m_Q~,~~~~~
{\bar\Lambda}=M_{\Lambda_Q}-m_Q~.
\end{equation}
\par
The main point in QCD sum rules for the IW function is to study the analytic
properties of the 3-point correlators,
\begin{mathletters}\label{3-correlator}
\begin{eqnarray}
\Xi_{\frac{1}{2}}(\omega,\omega',y)
&=&i^2\int d^4x d^4z e^{i(k^\prime\cdot x-k\cdot z)}
 \langle 0|{\cal T}j^{v^\prime}_{1/2,-}(x)\hbarc(0)\Gamma\hb(0)
 {\bar j}^v_{1/2,+}(z)|0\rangle\nonumber\\
&=&\frac{\Xi_{\rm hadron}(\omega,\omega',y)}{\sqrt{3}}
 \Big[\xi v^\alpha v^{\prime\alpha}\Omega^{(c\Gamma)}_{\alpha\beta}
   -\epsilon_c\Big(\tau_1^{(c)}v^\alpha v^\beta
     +\tau_2^{(c)}v^\alpha v^{\prime\beta}
     +\tau_3^{(c)}g^{\alpha\beta}\Big)\Omega^{(c\Gamma)}_{\alpha\beta}
  \nonumber\\
  && +\epsilon_b\Big(\tau_1^{(b)}v^\alpha v^\beta
     +\tau_2^{(b)}v^\alpha v^{\prime\beta}
     +\tau_3^{(b)}g^{\alpha\beta}\Big)\Omega^{(b\Gamma)}_{\alpha\beta}
 \Big]~,\\
\Xi_{\frac{3}{2}}^\mu(\omega,\omega',y)
&=&i^2\int d^4x d^4z e^{i(k^\prime\cdot x-k\cdot z)}
 \langle 0|{\cal T}j^{v^\prime\alpha}_{3/2,-}(x)\hbarc(0)\Gamma\hb(0)
 {\bar j}^v_{3/2,+}(z)|0\rangle\nonumber\\
&=&\Xi_{\rm hadron}(\omega,\omega',y)\Lambda_+^{\mu\alpha}
\Big[\xi v_\alpha \Gamma
   -\epsilon_c\Big(\tau_1^{(c)}v_\alpha v_\beta
     +\tau_2^{(c)}v_\alpha v^{\prime}_\beta
     +\tau_3^{(c)}g_{\alpha\beta}\Big)\gamma^\beta\Gamma\nonumber\\
  && +\epsilon_b\Big(\tau_1^{(b)}v_\alpha v_\beta
     +\tau_2^{(b)}v_\alpha v^{\prime}_\beta
     +\tau_3^{(b)}g_{\alpha\beta}\Big)\Gamma\gamma^\beta
 \Big]\Bigg(\Pa\Bigg)~.
\end{eqnarray}
\end{mathletters}
The variables $k$, 
$k'$ denote residual ``off-shell" momenta which are related to the momenta 
$P$ of the heavy quark in the initial state and $P'$ in the final state by 
$k=P-m_Qv$, $k'=P'-m_{Q'}v'$, respectively.
\par
The coefficient $\Xi(\omega,\omega',y)_{\rm hadron}$ in Eq.\ (\ref{3-correlator}) 
is an analytic function in the ``off-shell energies" $\omega=v\cdot k$ and 
$\omega'=v'\cdot k'$ with discontinuities for positive values of these 
variables. It furthermore depends on the velocity transfer $y=v\cdot v'$, 
which is fixed at its physical region for the process under consideration. By 
saturating with physical intermediate states in HQET, one finds the hadronic 
representation of the correlators  as following
\begin{equation}
\Xi_{\rm hadron}(\omega,\omega',y)
 =\frac{f_{1/2}f^*_{\Lambda_b}}
    {({\bar\Lambda}^\prime-\omega^\prime)({\bar\Lambda}-\omega)}
    +\mbox{higher resonances} \;.
\label{hadronic}
\end{equation}
In obtaining the above expression  the Dirac and Rartia-Schwinger spinor sums
\begin{eqnarray}
&& \Lambda_+=\sum_{s=1}^2u(v,s)\bar u(v,s)=\Pa~,\nonumber\\
&& \Lambda_{+}^{\mu\nu}=\sum_{s=1}^4u^\mu(v,s)\bar u^\nu(v,s)=
\Bigg(-g_t^{\mu\nu}+\frac{1}{3}\gamma_t^\mu\gamma_t^\nu\Bigg)\Pa~,
\end{eqnarray}
have been used, where $g^{\mu\nu}_t=g^{\mu\nu}-v^\mu v^\nu$. 
\par
\begin{figure}
\vskip 2cm
\begin{center}
\epsfig{file=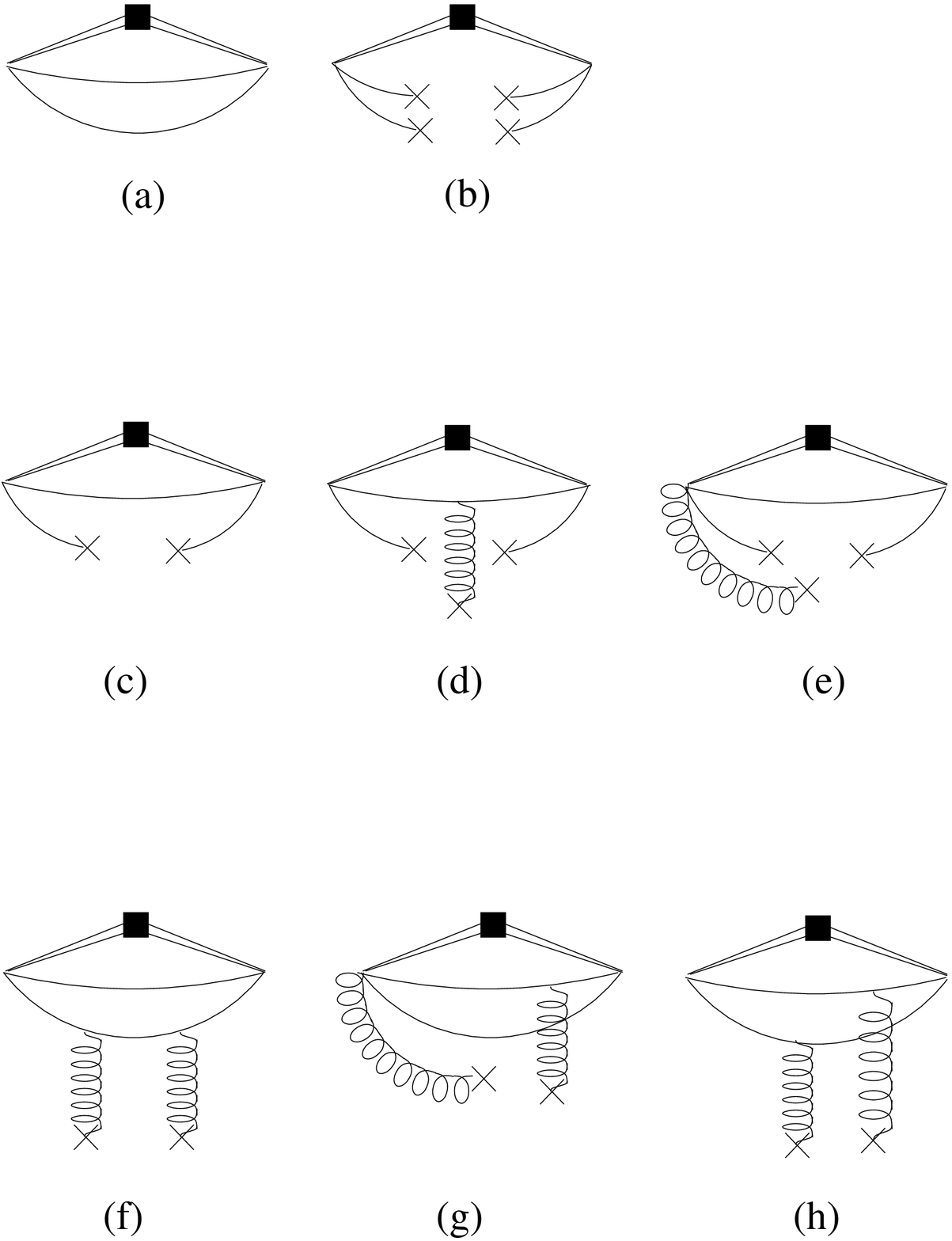, height=15cm}
\end{center}
\caption{}
\label{diagram}
\end{figure}

In the quark-gluon language, $\Xi(\omega,\omega',y)_{\frac{1}{2},\frac{3}{2}}$ 
in Eq.\ (\ref{3-correlator}) is written as
\begin{equation}
\Xi(\omega,\omega',y)_{\frac{1}{2},\frac{3}{2}}=\int^\infty_0 d\nu d\nu^\prime 
\frac{\rho^{\rm pert}(\nu,\nu^\prime,y)}
 {(\nu-\omega)(\nu^\prime-\omega^\prime)}+({\rm subtraction})
 +\Xi^{\rm cond}(\omega,\omega',y)\;,
\label{quarkgluon}
\end{equation}
where the perturbative spectral density function 
$\rho^{\rm pert}(\nu,\nu^\prime,y)$ and the condensate contribution 
$\Xi^{\rm cond}$ are related to the  calculation of the Feynman diagrams 
depicted in Fig.\ \ref{diagram} .
In Eq.\ (\ref{quarkgluon}), the $\gamma$-structures of spin-1/2 and 3/2 are
the same as those in Eq.\ (\ref{3-correlator}), respectively.
Subleading IW functions, $\tau_i^{(Q)}$, obtained from spin-1/2 and 3/2 are
therefore identical.
\par
The six $\tau_i^{(Q)}$ ($Q=c,b,~ i=1,2,3$) are not independent.
From the fact that
\begin{equation}
i\partial_\alpha(\hbarc\Gamma\hb)=
\hbarc(i\Dl_\alpha\Gamma+\Gamma i\Dr_\alpha)\hb=
({\bar\Lambda}v_\alpha-{\bar\Lambda'}v'_\alpha)\hbarc\Gamma\hb~,
\end{equation}
Eq.\ (\ref{subleading}) implies
\begin{equation}
(\tau_1^{(c)}+\tau_1^{(b)})v_\alpha v_\beta
+(\tau_2^{(c)}+\tau_2^{(b)})v_\alpha v'_\beta
+(\tau_3^{(c)}+\tau_3^{(b)})g_{\alpha\beta}
=({\bar\Lambda} v_\beta-{\bar\Lambda'}v'_\beta)v_\alpha\xi(y)~.
\end{equation}
The above expression relates $\tau_i^{(c)}$ with $\tau_i^{(b)}$ as
\begin{mathletters}\label{taucb}
\begin{eqnarray}
\tau_1^{(c)}+\tau_1^{(b)}&=&{\bar\Lambda}\xi~,\\
\tau_2^{(c)}+\tau_2^{(b)}&=&-{\bar\Lambda'}\xi~,\\
\tau_3^{(c)}+\tau_3^{(b)}&=&0~.
\end{eqnarray}
\end{mathletters}
Other relations are obtained from the equation of motion of the heavy quark,
$v\cdot D h^{(Q)}_v=0$:
\begin{mathletters}\label{eom}
\begin{eqnarray}
\hbarc iv\cdot\Dl\Gamma\hb&=&{\bar\psi}^\alpha_{v'}\Big(
 y\tau_1^{(c)}+\tau_2^{(c)}\Big)\Gamma\Lambda_v=0~,\\
\hbarc\Gamma iv\cdot \Dr\hb&=&{\bar\psi}^\alpha_{v'}\Gamma\Big(
 \tau_1^{(b)}+y\tau_2^{(b)}+\tau_c^{(b)}\Big)\Lambda_v=0~,
\end{eqnarray}
\end{mathletters}
From the above 5 equations in Eq.\ (\ref{taucb}), (\ref{eom}), all the six
subleading IW functions are reduced to only one independent form factor.
We just pick up $\tau_1^{(b)}(y)\equiv\tau(y)$, then others are
\begin{mathletters}
\label{sixtau}
\begin{eqnarray}
\tau_1^{(c)}&=&{\bar\Lambda}\xi-\tau~,\\
\tau_2^{(c)}&=&-y{\bar\Lambda}\xi+y\tau~,\\
\tau_3^{(c)}&=&y(y{\bar\Lambda}-{\bar\Lambda'})\xi-(y^2-1)\tau~,\\
\tau_2^{(b)}&=&(y{\bar\Lambda}-{\bar\Lambda'})\xi-y\tau~,\\
\tau_3^{(b)}&=&-y(y{\bar\Lambda}-{\bar\Lambda'})\xi+(y^2-1)\tau~,
\end{eqnarray}
\end{mathletters}
\par
Now that all the subleading IW functions are related to $\tau(y)$, we have only
to extract the coefficient of $v^\alpha v^\beta\Omega^{(b\Gamma)}_{\alpha\beta}$
(or $\Lambda^{\mu\alpha}_+v_\alpha v_\beta\Gamma\gamma^\beta$ for spin 3/2)
in Eqs.\ (\ref{3-correlator}) and (\ref{quarkgluon}).
\par
The QCD sum rule is obtained by equating the phenomenological and 
theoretical expressions for $\Xi$. In doing this the quark-hadron duality 
needs to be assumed to model the contributions of higher resonance part of 
Eq. (\ref{hadronic}). Generally speaking, the duality is to simulate the 
resonance contribution by the perturbative part above some thresholds 
$\omega_c$ and $\omega'_c$, that is
\begin{equation}
{\rm res.}=\int^\infty_{\omega_c}\int^\infty_{\omega^\prime_c}d\nu 
 d\nu^\prime~\frac{\rho^{\rm pert}(\nu,\nu^\prime,y)}
 {(\nu-\omega)(\nu^\prime-\omega^\prime)}~.
\label{res}
\end{equation} 
In the QCD sum rule analysis for  $B$ semileptonic decays into ground state 
$D$ mesons, it was argued by Neubert in \cite{Neubert}, and Blok and Shifman 
in \cite{Shifman} that the perturbative and the hadronic spectral densities 
can not be locally dual to each other, and therefore the necessary way to 
restore duality 
is to integrate the spectral densities over the ``off-diagonal'' variable 
$\nu_-=\sqrt{\frac{y+1}{y-1}}(\nu-\nu')/2$, keeping the ``diagonal'' variable 
$\nu_+=(\nu+\nu')/2$ fixed. It is in $\nu_+$ that the quark-hadron duality is 
assumed for the integrated spectral densities.  The same prescription shall be 
adopted in the following analysis.  On the other hand, in order to suppress the 
contributions of higher resonance states a double Borel transformation in 
$\omega$ and $\omega'$ is performed to both sides of the sum rule, which 
introduces two Borel parameters $T_1$ and $T_2$.
\par
Combining Eqs. (\ref{hadronic}), (\ref{quarkgluon}), our duality assumption 
and making the double Borel transformation, one obtains the sum rule for 
$\xi(y)$ as follows;
\begin{eqnarray}\label{doubleBT}
\lefteqn{
Mf_{1/2}f^*_{\Lambda_b}e^{-{\bar\Lambda}^\prime/2T'}e^{-{\bar\Lambda}/2T}
 \Bigg(\Pb\Bigg)C_\Gamma\Bigg(\Pa\Bigg)}\nonumber\\
&=&
2\;\Bigg(\frac{y-1}{y+1}\Bigg)^{1/2}
\int^{\omega_c(y)}_0 d\nu_+ \int^{\nu_+}_{-\nu_+}d\nu_-
\exp\Bigg(-\frac{\nu_+-\sqrt{\frac{y-1}{y+1}}\nu_-}{2T'}
 -\frac{\nu_++\sqrt{\frac{y-1}{y+1}}\nu_-}{2T}\Bigg)
\rho(\nu_+,\nu_-;y) \nonumber\\
&&+{\hat B}^{\omega^\prime}_{2T'}{\hat B}^\omega_{2T}\Xi^{\rm cond}~,
\label{sumrule}
\end{eqnarray}
where $\nu=\nu_++\sqrt{\frac{y-1}{y+1}}\nu_-$, 
$\nu^\prime=\nu_+-\sqrt{\frac{y-1}{y+1}}\nu_-$, and
\label{C}
\begin{eqnarray}
C_\Gamma=\left\{\begin{array}{l}
\frac{1}{\sqrt{3}}\Bigg[
\xi v^\alpha v^{\prime\alpha}\Omega^{(c\Gamma)}_{\alpha\beta}
   -\epsilon_c\Big(\tau_1^{(c)}v^\alpha v^\beta
     +\tau_2^{(c)}v^\alpha v^{\prime\beta}
     +\tau_3^{(c)}g^{\alpha\beta}\Big)\Omega^{(c\Gamma)}_{\alpha\beta}\\
   +\epsilon_b\Big(\tau_1^{(b)}v^\alpha v^\beta
     +\tau_2^{(b)}v^\alpha v^{\prime\beta}
     +\tau_3^{(b)}g^{\alpha\beta}\Big)\Omega^{(b\Gamma)}_{\alpha\beta}
\Bigg]~~~~~({\rm for~spin~1/2})~,\\
\xi v_\alpha \Gamma
   -\epsilon_c\Big(\tau_1^{(c)}v_\alpha v_\beta
     +\tau_2^{(c)}v_\alpha v^{\prime}_\beta
     +\tau_3^{(c)}g_{\alpha\beta}\Big)\gamma^\beta\Gamma\\
   +\epsilon_b\Big(\tau_1^{(b)}v_\alpha v_\beta
     +\tau_2^{(b)}v_\alpha v^{\prime}_\beta
     +\tau_3^{(b)}g_{\alpha\beta}\Big)\Gamma\gamma^\beta
~~~~~~({\rm for~spin~3/2})~.
\end{array}\right.
\end{eqnarray}
\par
Now the remaining thing is to evaluate the relevant diagrams in 
Fig.\ \ref{diagram}.
The leading contributions are given in \cite{PLB502}.
For the subleading corrections to the perturbative spectral density function
$\rho(\omega,\omega';y)$, we have
\label{pert}
\begin{eqnarray}
\rho(\omega,\omega';y)&=&
{\hat B}^{-z'}_{1/\omega'}{\hat B}^{-z}_{1/\omega}
{\hat B}^{\omega'}_{1/z'}{\hat B}^{\omega}_{1/z}\Xi^{\rm pert}\nonumber\\
&=&\Bigg(\frac{6N_c! ai}{\pi^4}\Bigg)\Omega_{\alpha\beta}
\frac{1}{2\sinh^7\theta}\Theta(\omega)\Theta(\omega')
 \Theta(2y\omega'\omega-\omega^2-\omega'^2) \nonumber\\
&&
\Bigg[
 \frac{2v^\alpha v^{'\beta}}{\sinh^2\theta} \Bigg(
   \frac{2\cosh\theta A^3B^3}{3!3!}-\frac{e^{-\theta}A^2B^4}{2!4!}
   -\frac{e^\theta A^4B^2}{4!2!}\Bigg)\nonumber\\
&&
+\frac{2v^\alpha v^\beta}{\sinh^2\theta}\Bigg(
  \frac{e^{2\theta}A^4B^2}{4!2!}+\frac{e^{-2\theta}A^2B^4}{2!4!}
  -\frac{2A^3B^3}{3!3!}\Bigg)
-g^{\alpha\beta}\frac{A^3B^3}{3!3!}
\Bigg]~,
\end{eqnarray}
from the perturbative diagram Fig.\ \ref{diagram} (a), where
\begin{mathletters}
\begin{eqnarray}
\Omega_{\alpha\beta}&\equiv&-i\epsilon_c\Omega^{(c\Gamma)}_{\alpha\beta}
 +i\epsilon_b\Omega^{(b\Gamma)}_{\alpha\beta}~,\\
A&\equiv&\omega'-\omega e^{-\theta}~,~~~~~
B\equiv\omega e^\theta-\omega'~,\\
e^\theta&\equiv&y+\sqrt{y^2-1}~.
\end{eqnarray}
\end{mathletters}
\par
For the condensate contributions we just give results when $T'=T$ for 
simplicity;
\begin{mathletters}\label{cond}
\begin{eqnarray}
{\hat B}^{\omega'}_{2T}{\hat B}^\omega_{2T}\Xi^{\langle{\bar q}q\rangle}&=&
-\frac{ibg^{\alpha\beta}\Omega_{\alpha\beta}}{2\pi^2(1+y)^2}\Bigg[
 64\qq T^5-\frac{1}{3}\qGq T^3(4y+5/2)\Bigg]\nonumber\\
&&
-\frac{ib v^\alpha\Omega_{\alpha\beta}}{4\pi^2(1+y)^3}\Bigg[
 -128\qq T^5(3v+2v')^\beta \nonumber\\
&&
   +\frac{4}{3}\qGq \Big\{
   (6y+7/2)v^\beta+(y-3/2)v'^\beta\Big\}\Bigg]~,\\
{\hat B}^{\omega'}_{2T}{\hat B}^\omega_{2T}\Xi^{\qGq}&=&
-\frac{ib\qGq T^3}{12(1+y)^3}\Omega_{\alpha\beta}\Big[
 -2g^{\alpha\beta}(2y^2+3y+1)
 +(10y+6)v^\alpha v^\beta+4y v^\alpha v'^\beta\Big]~,\\
{\hat B}^{\omega'}_{2T}{\hat B}^\omega_{2T}\Xi^{\GG}&=&
\frac{ia\GG T^4}{192\pi^3(1+y)^5}\Omega_{\alpha\beta}\Big[
 8(y+1)^2(y-2)\big\{-g^{\alpha\beta}+5v^\alpha(v+v')^\beta\big\}\nonumber\\
&&
 +24(y-1)v^\alpha v'^\beta-16(y+1)(y+4)v^\alpha v^\beta\Big]\nonumber\\
&&
-\frac{ia\GG T^4}{512\pi^3(1+y)^4}\Omega_{\alpha\beta}\Big[
  -2(1+y)g^{\alpha\beta}+6v^\alpha(v+v')^\beta\Big]~.
\end{eqnarray}
\end{mathletters}
Note that these results are from $\Lca$.
If $\Lcb$ were the final state, $\Omega_{\alpha\beta}$ would be replace by 
a proper $\gamma$-structure, leaving all the other things unchanged.

\section{Results and Discussions}

For the numerical analysis, the standard values of the condensates are used;
\begin{eqnarray}
\langle{\bar q}q\rangle&=&-(0.23~{\rm GeV})^3~,\nonumber\\
\langle\alpha GG\rangle&=&0.04~{\rm GeV}^4~,\nonumber\\
\langle \bar{q}g\sigma\cdot G q\rangle&\equiv&m_0^2\langle{\bar q}q
\rangle~,~~~~~m_0^2=0.8~{\rm GeV}^2~.
\end{eqnarray}
There are many parameters engaged in the QCD sum rule calculations.
The key point in the numerical analysis is to find a reasonable parameter space 
where the QCD sum rule results are stable.
First, the continuum threshold $\omega_c^\prime$ in 
$f_{\frac{1}{2}(\frac{3}{2})}$
(${\bar\Lambda}^\prime$) can differ from that in $f_{\Lambda_b}$ 
(${\bar\Lambda}$).  
However, it is expected that the values of $\omega_c$ and 
$\omega_c'$ would not be different significantly.
This is because the mass 
difference ${\bar\Lambda}^\prime-{\bar\Lambda}$ is fairly small \cite{PLB476},
${\bar\Lambda}^\prime-{\bar\Lambda}\simeq 0.2~{\rm GeV}$.  
Indeed the central 
values of them were close to each other in the sum rules analysis for 
$f_{\frac{1}{2}(\frac{3}{2})}$ (${\bar\Lambda}^\prime$) and $f_{\Lambda_b}$ 
(${\bar\Lambda}$).
One more thing to be noticed here is that the continuum threshold 
$\omega_c$ in Eq. (\ref{sumrule}) can be a function of $y$ in general.
But for simplicity, we take it to be a 
constant $\omega_c(y)=\omega_c=\omega_c'=\omega_0$ in the numerical analysis.  
In this sense, we use only one constant continuum threshold throughout the 
analysis.
An alternative choice of $\omega_c(y)=(1+y)\omega_0/2y$ is suggested in 
Ref.\ \cite{Neubert}.
We find that this choice yields almost no numerical differences.
This is because the kinematically allowed region is very narrow around the
zero recoil.
\par
Second, there are input parameters of $a$ and $b$ in the interpolating fields
in Eq.\ (\ref{deriv}).  
They are the parameters that generalize pseudoscalar or axial-vector nature
of the light degrees of freedom ($\Gamma_{1/2,3/2}$ in Eq.\ \ref{deriv}).
In Ref.\ \cite{PLB476}, a particular choice of $(a,b)=(1,0)$ gives the best 
stability for the mass parameter ${\bar\Lambda}^\prime$.  
We adopt the same choice of $(a,b)=(1,0)$ in the present analysis.  
\par
Third, there are two Borel parameters $T_1$ and $T_2$ distinct in general, 
corresponding to $\omega$ and $\omega^\prime$ in 
$\Xi(\omega, \omega',y)$, respectively.  
We have taken $T_1=T_2$ in the analysis.  
In Ref.\ \cite{Huang} for $B$ into excited charmed meson transition,
the authors found a $10\%$ increase in the leading IW function at zero recoil 
when $T_2/T_1=1.5$ as compared to the case when $T_1=T_2$.  
It seems quite reasonable for one to expect that in the case of heavy baryon, 
the numerical results should be 
similar for the small variations around $T_2/T_1=1$.
\par
In short, we adopt the same parameters used in \cite{PLB476,PLB502} where
the mass parameter and the leading IW function are calculated.
It makes sense because the observables involved are directly related to the
subleading IW function $\tau(y)$ through Eq.\ (\ref{doubleBT}).
\par
\begin{figure}
\begin{center}
\epsfig{file=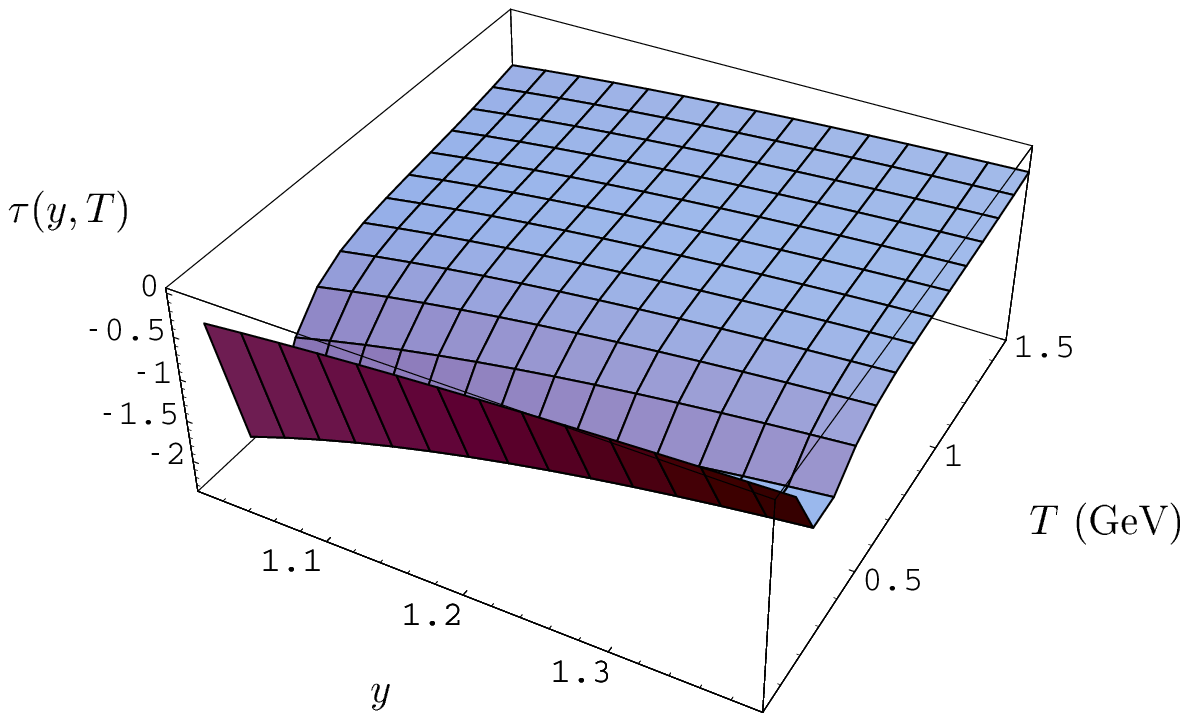}
\end{center}
\caption{}
\label{3d}
\end{figure}
\begin{figure}
\begin{center}
\epsfig{file=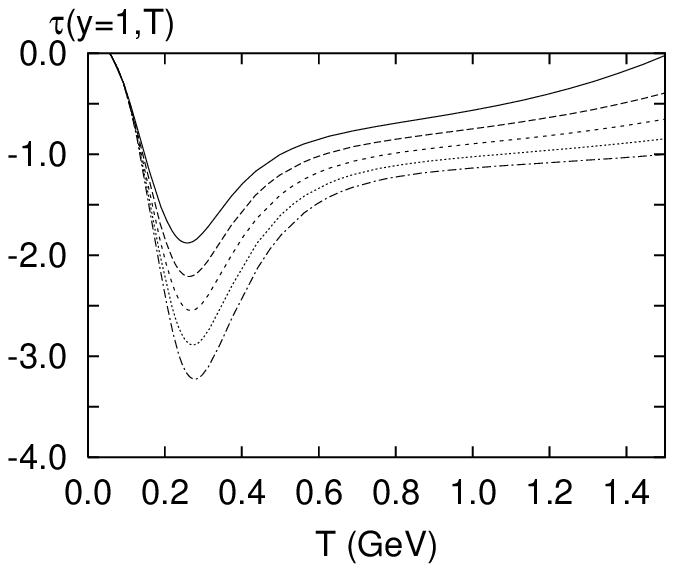}
\end{center}
\caption{}
\label{tauT}
\end{figure}
\begin{figure}
\begin{center}
\epsfig{file=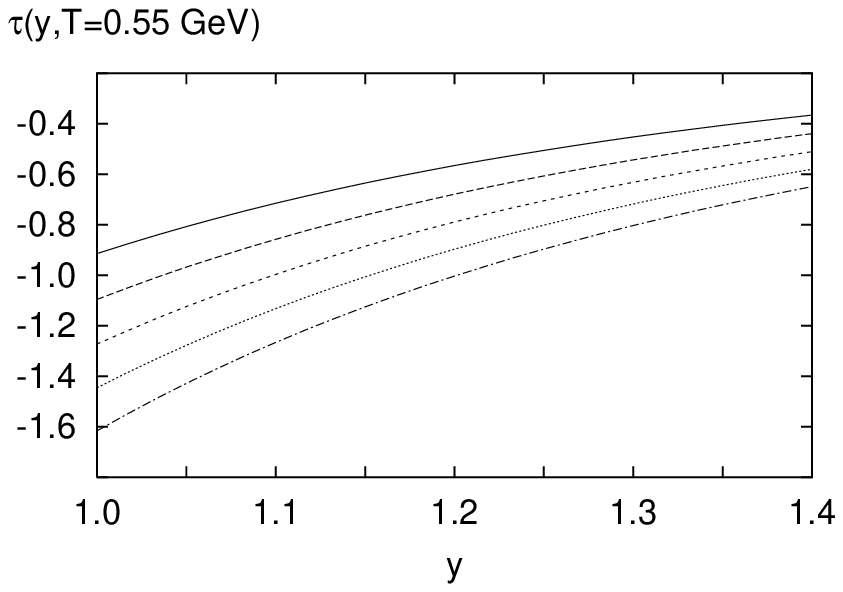}
\end{center}
\caption{}
\label{tauy}
\end{figure}

In Fig.\ \ref{3d}, $\tau$ is plotted as a function of $(y,T)$.
Figure \ref{tauT} shows the stability of $\tau(y=1)$ for the Borel parameter.
The sum rule window is 
\begin{equation}
0.1 \le T\le 1.0~({\rm GeV})~.
\label{window}
\end{equation}
The upper and lower bounds are fixed such that the pole contribution amounts to
50\% while the condensate one to 12\%.
One notes that the window given in Eq.\ (\ref{window}) overlaps those obtained
in the Refs.\ \cite{sumrule,PLB476,PLB502}.
Of course, this reflects the self-consistency of the sum rule analysis.
In Fig.\ \ref{tauy}, we present the shape of $\tau(y)$ for a fixed Borel 
parameter.
We found that
\begin{eqnarray}
\tau(y)&=&\tau(1)[1-\rho^2(y-1)],\nonumber\\
\tau(1)&=&-1.27^{-0.17}_{+0.18}~{\rm GeV},~~~
   {\rm for}~\omega_0=1.4\pm 0.1~{\rm GeV}~,\nonumber\\
\rho^2&=&2.76^{-0.004}_{+0.008}~,~~~{\rm for}~\omega_0=1.4\pm 0.1~{\rm GeV}~~.
\end{eqnarray}

\section{Summary}

Subleading contributions of ${\cal O}(1/m_Q)$ to the $\Lb\to\Lce$ weak form 
factors are important because some of the form factors do not survive at the 
heavy quark limit, and other remaining form factors vanish at zero recoil.
Using the QCD sum rules, we calculate the subleading IW function $\tau(y)$ 
which appears in the current matching in the HQET at ${\cal O}(1/m_Q)$.
We obtain $\tau(y)$ given by
\begin{equation}
\tau(y)=-1.27[1-2.76(y-1)]~{\rm GeV}~.\nonumber\\
\end{equation}
The best stability is attained when the continuum threshold $\omega_0=1.4$ GeV.
The parameter space for the analysis is the same as previous one for the leading
IW function.
The fact that by using the same set of parameters the present sum rule window
for the mass parameter, leading and NLO IW function overlaps the previous ones
ensures the self-consistenncy of the QCD sum rules.
Our results can be applied directly to the decay mode $\Lb\to\Lce\ell\bar{\nu}$,
along with the use of the previous LO IW function, but a complete analysis at
${\cal O}(1/m_Q)$ requires the information on another NLO contributions from the
HQET Lagrangian.

\begin{center}
{\large\bf Acknowledgements}\\[10mm]
\end{center}
This work was supported by the BK21 Program of the Korean Ministry of Education.
The work of GTP was supported in part by Yonsei University Research Fund of 
2000.
\begin{appendix}
\section{Contractions of $\Omega_{\alpha\beta}$}

After a simple algebra, possible contractions for $\Omega_{\alpha\beta}$
are given by
\begin{eqnarray}
v^{\prime\alpha}\Omega_{\alpha\beta}&=&0~,\nonumber\\
v^\alpha v^\beta\Omega_{\alpha\beta}^{(cV)}&=&\Bigg(\Pb\Bigg)\Big[
   -2yv^\mu\gamma_5\Big]\Bigg(\Pa\Bigg)~,\nonumber\\
v^\alpha v^{\prime\beta}\Omega_{\alpha\beta}^{(cV)}&=&\Bigg(\Pb\Bigg)\Big[
   (y-1)\gamma^\mu\gamma_5-2v^\mu\gamma_5\Big]\Bigg(\Pa\Bigg)~,\nonumber\\
g^{\alpha\beta}\Omega_{\alpha\beta}^{(cV)}&=&\Bigg(\Pb\Bigg)\Big[
   3\gamma^\mu\gamma_5\Big]\Bigg(\Pa\Bigg)~,\nonumber\\
v^\alpha v^\beta\Omega_{\alpha\beta}^{(cA)}&=&\Bigg(\Pb\Bigg)\Big[
   -2yv^\mu+2\gamma^\mu\Big]\Bigg(\Pa\Bigg)~,\nonumber\\
v^\alpha v^{\prime\beta}\Omega_{\alpha\beta}^{(cA)}&=&\Bigg(\Pb\Bigg)\Big[
   (y+1)\gamma^\mu-2v^\mu\Big]\Bigg(\Pa\Bigg)~,\nonumber\\
g^{\alpha\beta}\Omega_{\alpha\beta}^{(cA)}&=&\Bigg(\Pb\Bigg)\Big[
   3\gamma^\mu\Big]\Bigg(\Pa\Bigg)~,\nonumber\\
v^\alpha v^\beta\Omega_{\alpha\beta}^{(bV)}&=&\Bigg(\Pb\Bigg)\Big[
   (y-1)\gamma^\mu\gamma_5-2v^\mu\gamma_5\Big]\Bigg(\Pa\Bigg)~,\nonumber\\
v^\alpha v^{\prime\beta}\Omega_{\alpha\beta}^{(bV)}&=&\Bigg(\Pb\Bigg)\Big[
   (1-y)\gamma^\mu\gamma_5+2v^\mu\gamma_5-2(y+1)v^{\prime\mu}\gamma_5
    \Big]\Bigg(\Pa\Bigg)~,\nonumber\\
g^{\alpha\beta}\Omega_{\alpha\beta}^{(bV)}&=&\Bigg(\Pb\Bigg)\Big[
   -\gamma^\mu\gamma_5-2v^{\prime\mu}\gamma_5\Big]\Bigg(\Pa\Bigg)~,\nonumber\\
v^\alpha v^\beta\Omega_{\alpha\beta}^{(bA)}&=&\Bigg(\Pb\Bigg)\Big[
   (y+1)\gamma^\mu-2v^\mu\Big]\Bigg(\Pa\Bigg)~,\nonumber\\
v^\alpha v^{\prime\beta}\Omega_{\alpha\beta}^{(bA)}&=&\Bigg(\Pb\Bigg)\Big[
   (y+1)\gamma^\mu-2v^\mu+2(y-1)v^{\prime\mu}\Big]\Bigg(\Pa\Bigg)~,\nonumber\\
g^{\alpha\beta}\Omega_{\alpha\beta}^{(bA)}&=&\Bigg(\Pb\Bigg)\Big[
\gamma^\mu+2v^{\prime\mu}\Big]\Bigg(\Pa\Bigg)~,
\label{contOmega}
\end{eqnarray}
where $V(A)\equiv\gamma^\mu(\gamma^\mu\gamma_5)$.

\end{appendix}


\newpage

\begin{center}{\large\bf FIGURE CAPTIONS}\end{center}

\noindent
Fig.~1
\\
Feynman diagrams for the three-point function with derivative 
interpolating fields.  Double line denotes the heavy quark.
\vskip .3cm
\par

\noindent
Fig.~2
\\
Three dimensional plot of $\tau$ as a function of $y$ and $T$ in units of GeV.
The continuum threshold is chosen to be $\omega_c(y)=1.4$ GeV.
\vskip .3cm
\par

\noindent
Fig.~3
\\
$\tau(1)$ as a function of the Borel parameter $T$.
Each graph corresponds to $\omega_0=1.2,1.3,1.4,1.5,1.6$ GeV, respectively, 
from the top.
\vskip .3cm
\par

\noindent
Fig.~4
\\
$\tau(y)$ at a fixed Borel parameter $T=0.34$.
Each graph corresponds to $\omega_0=1.2,1.3,1.4,1.5,1.6$ GeV, respectively, 
from the top.
\vskip .3cm
\par

\pagebreak


\pagebreak


\newpage

\end{document}